\documentclass[10pt,journal,compsoc]{IEEEtran}

\usepackage{booktabs}
\usepackage{color}

\usepackage{enumitem}
\usepackage{ifpdf}

\usepackage{multirow}
\usepackage[table,xcdraw]{xcolor}

\usepackage{amsmath}

\usepackage[switch,pagewise]{lineno} 

\usepackage{acronym}
\acrodef{VR}[VR]{virtual reality}
\acrodef{AR}[AR]{augmented reality}
\acrodef{MR}[MR]{mixed reality}
\acrodef{DR}[DR]{diminished reality}
\acrodef{SWI}[SWI]{size-weight illusion}
\acrodef{FOV}[FOV]{field of view}
\acrodef{HMD}[HMD]{head-mounted display}
\acrodef{ANOVA}[ANOVA]{analysis of variance}
\acrodef{DOF}[DOF]{degrees of freedom}
\acrodef{SD}[SD]{standard deviation}
\acrodef{COG}[COG]{center of gravity}

\ifCLASSOPTIONcompsoc
  \usepackage[nocompress]{cite}
\else
  \usepackage{cite}
\fi

\ifCLASSINFOpdf
  \usepackage[pdftex]{graphicx}
  \usepackage{orcidlink}

  \DeclareGraphicsExtensions{.pdf,.jpeg,.png}
\else
\fi

\begin{document}
\title{Perceived Weight of Mediated Reality Sticks}

\author{Satoshi~Hashiguchi*~\orcidlink{0009-0002-0427-030X},
        Yuta~Kataoka*~\orcidlink{0000-0001-6144-6477},
        Asako~Kimura~\orcidlink{0000-0002-9859-9079},
        and~Shohei Mori~\orcidlink{0000-0003-0540-7312},~\IEEEmembership{Member,~IEEE}
\IEEEcompsocitemizethanks{
    \IEEEcompsocthanksitem S. Hashiguchi, Y. Kataoka, and A. Kimura are with Ritsumeikan University, Japan. \protect
    \IEEEcompsocthanksitem S. Mori is with the Visualization Research Center (VISUS), Cluster of Excellence IntCDC, University of Stuttgart, Germany. \protect
    \IEEEcompsocthanksitem (*) for equal contribution. \protect
}
\thanks{Manuscript received March 31, 2024; revised June 14, 2025; accepted July 3, 2025.}}

\markboth{© 2025 IEEE. Accepted for publication in IEEE TVCG. DOI: \url{https://doi.org/10.1109/TVCG.2025.3591181}}%
{Shell \MakeLowercase{\textit{et al.}}: Bare Demo of IEEEtran.cls for Computer Society Journals}

\IEEEtitleabstractindextext{%
\begin{abstract}
Mediated reality, where augmented reality (AR) and diminished reality (DR) meet, enables visual modifications to real-world objects. 
A physical object with a mediated reality visual change retains its original physical properties. However, it is perceived differently from the original when interacted with.
We present such a mediated reality object, a stick with different lengths or a stick with a missing portion in the middle, to investigate how users perceive its weight and center of gravity. We conducted two user studies ($N=10$), each of which consisted of two substudies. We found that the length of mediated reality sticks influences the perceived weight. A longer stick is perceived as lighter, and vice versa. The stick with a missing portion tends to be recognized as one continuous stick. Thus, its weight and center of gravity (COG) remain the same. We formulated the relationship between inertia based on the reported COG and perceived weight in the context of dynamic touch.
\end{abstract}

}

\maketitle

\IEEEdisplaynontitleabstractindextext
\IEEEpeerreviewmaketitle

\IEEEraisesectionheading{\section{Introduction}\label{sec:introduction}}

\IEEEPARstart{T}{he} human senses do not work individually; rather, they interact and influence each other in a phenomenon known as the ``cross-modal effect.'' This effect has attracted attention in experience design in \ac{MR} \cite{streit2007rotational, Lecuyer00}.
One example of such an effect is the \ac{SWI} \cite{charpentier1981experimental}, in which objects with the same mass are perceived as having different weights depending on their sizes and shapes.
\ac{MR} research has demonstrated that the visual alterations of a physical object can induce different weights in our perception.
Developing guidelines for the design space and clarifying relationships between modified appearances and perceived weights would enable better-controlled experiences with haptic feedback systems that have reduced hardware and form factors.

The literature has demonstrated that illusory experiences can occur within a wide range of \ac{MR} paradigms \cite{issartel2015perceiving}.
\Ac{VR} has become a popular tool for demonstrating these effects because of the availability of affordable hardware and software \cite{Zenner17,Fujinawa17}. However, the impact of visual fidelity and personalization (e.g., personalized limbs) on perceptions within \ac{VR} environments is still a topic of debate.
\Ac{AR} addresses some of these issues by incorporating the real environment, either optically or through video passthrough systems \cite{Schmalstieg2016}.

\begin{figure}[!t]
    \centering
    \includegraphics[width=\columnwidth]{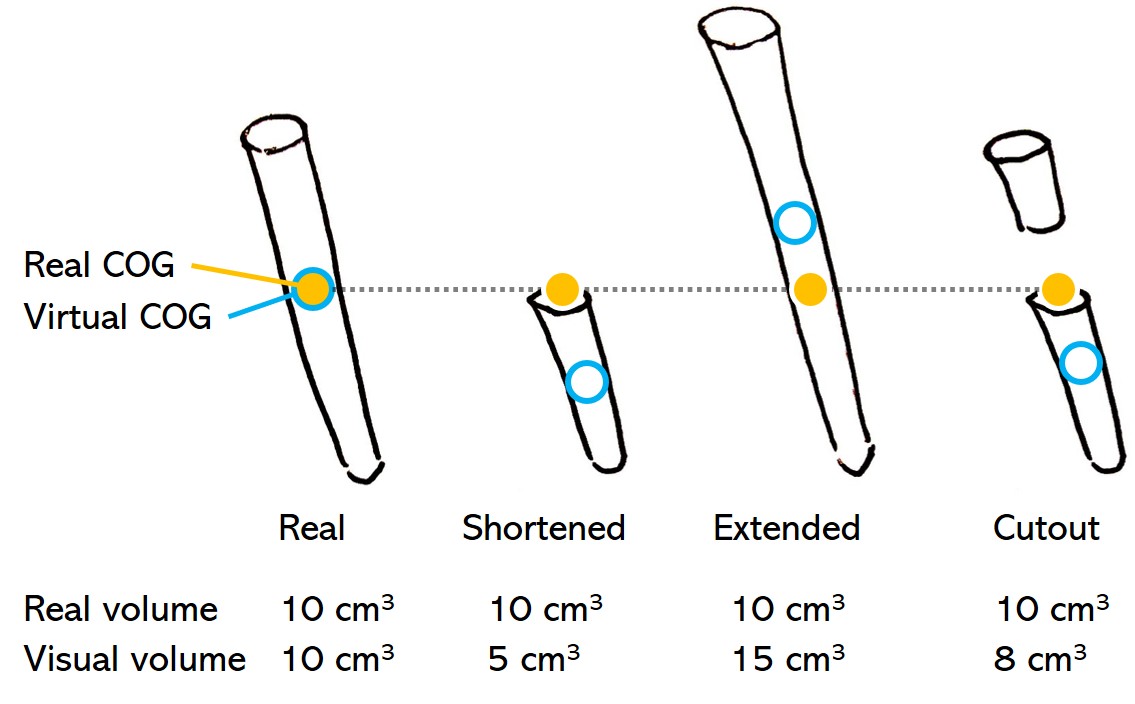}
    \caption{
    Mediated reality sticks with weight--COG
inconsistencies. All illustrated sticks have the same physical weights and COGs but different mediated realities. We investigate the characteristics of mediated reality sticks in different lengths in Experiment 1. Mediated reality sticks with cutout parts are analyzed in Experiment 2.
    Note that the cutout stick could have different COGs depending on whether the user recognizes the object as a whole or two separate objects.
    To grasp this understanding, besides measuring the perceived weight, we collected the perceived COG as an alternative length-related measure, since the length was apparent by the appearance.
    }
    \label{fig:teaser}
\end{figure}

However, achieving accurate visual augmentation in optical see-through AR can be technically challenging \cite{itoh2021towards}, and certain types of visual modifications are not conceptually feasible because of their overlay natures \cite{mori2017survey}. For example, it is impossible to make real objects appear smaller solely through the use of \ac{AR} technology. \Ac{DR} offers a means of removing objects from the \ac{AR} environment (i.e., mediated reality \cite{mann1999mediated,mori2017survey}), making visual modifications more flexible as in \ac{VR}.
In this study, we demonstrate a weight illusion caused by visual changes using both tools, where a real object or a stick is stretched in length or partly removed, here investigating the effect of visual changes on the perception of overall weight (\figurename~\ref{fig:teaser}).

Historically, MR research has explored SWI to investigate visual changes leading to different perceived weights. For quantification, we attempt to formulate correlations between perceived weights and a visually recognized physical property, namely, \ac{COG}, with the help of founded formulations in dynamic touch.

Our contributions are summarized as follows.
\begin{itemize}
    \item We demonstrate that virtually longer or shorter sticks (i.e., mediated reality sticks) can alter their perceived weights and \ac{COG} locations when actively perceived.
    \item We show that longer and shorter sticks are perceived as lighter and heavier, respectively.
    \item We show that perceived COG locations vary depending on the appearance of their lengths.
    \item We show that sticks with a missing part in their middles present consistent weights and COG locations, suggesting that the sticks are recognized as a whole.
     \item We further discuss how inertia based on the reported COG can explain the perceived weight.
\end{itemize}
Note that the present paper builds on our previous paper \cite{hashiguchi2018perceived}, with refined analysis and deeper discussions, as well as assessments of different visual stimuli (Cutout in \figurename~\ref{fig:teaser}).

\section{Related Work}
We review the characteristics of human haptic perception and the influence of visual information on it, such as colors and objects' dynamics, on haptic sensations in different reality paradigms.

\subsection{Real-World Illusions}

Humans can perceive objects' properties with physical interaction (e.g., grasping or manipulating hand-held objects \cite{gibson1966senses, loomis1986tactual}). Notably, according to our daily experiences, the accurate estimation of characteristics such as length \cite{solomon1988haptically}, weight \cite{ross1978eh}, and content \cite{plaisier2017many} can typically be accomplished even in the absence of visual cues.
For example, the length of a stick is often perceived by wielding it to explore a particular axis of rotation \cite{riley2002perceptual, stephen2010role}. The \ac{COG} location also influences the estimated length and weight of a stick \cite{solomon1988haptically, amazeen1996weight}.
Visual stimuli, nonetheless, impact the alteration of the perceived sensation derived from haptic stimuli. The perceived weight of an object is influenced by factors such as its size \cite{charpentier1981experimental}, shape \cite{dresslar1894studies}, material \cite{flanagan1995effects}, and brightness \cite{walker2010brightness}.

The SWI is a phenomenon of varying object weights depending on the object’s apparent size \cite{charpentier1981experimental}.
A larger object is perceived as lighter than a smaller object, even when both objects possess the same mass.
The SWI is utilized to induce changes in applied forces during object manipulation \cite{flanagan2000independence}. Factors such as the material and brightness of an object also contribute to the degree of the \ac{SWI} \cite{buckingham2013size, vicovaro2019influence}.

The underlying mechanism of the \ac{SWI} remains a subject of discussion. One possible explanation for the illusion lies in the discrepancy between the predicted and actual weights of the grasped object \cite{woodworth1924psychology}.
In general, human perception can be modeled based on sensory processes, where multiple sensory modalities from receptors are integrated for example, by the maximum likelihood estimation of the probability distribution of each sensory modality \cite{ernst2002humans}. However, these perceptual models were constructed within the context of specific experimental conditions and environments.

Previous studies have not yet elucidated the relationship between the visual alteration, neither extension nor reduction of a stick object, and the perception of its weight or \ac{COG}. We help provide insights into human perceptual properties by examining an object with unique visual effects (\figurename~\ref{fig:teaser}).
We formulate relationships between the perceived weight and inertia. We especially discuss inertia based on the reported COG better explains the perceived weight compared to the analytic COG.

\subsection{Illusions in Virtual Reality}
VR can present arbitrary environments with reasonably simulated real-world phenomena.
This key feature of VR allows researchers to simulate multimodal stimuli that are difficult to replicate in reality, thus advancing psychological studies, especially in visuo-haptics.
There are successful VR examples of altering the perceived weight of a passive object with a motorized spindle \cite{Zenner17},
changing the heaviness of a virtual object using tracking offsets \cite{rietzler2018breaking}, and changing the softness of a chair by shifting the user's viewpoint \cite{matsumuro2023modified}.

However, as mentioned in Section \ref{sec:introduction}, real--virtual discrepancies because of the technical limitations of real-time high-fidelity graphics \cite{lonne2023effect}, haptics \cite{ogawa2018object}, and other modalities \cite{kurihara2014haptic}, necessitate real-to-virtual mapping, which is still imperfect.
These discrepancies become more evident especially in visuo-haptic scenarios, because users need to touch with reasonably well-simulated or virtual objects through their avatars' bodies, which induces a sense of ownership \cite{ujitoko2021survey}.
To reduce unnecessarily arguable concerns, we rely on mediated reality technology, which preserves the real-world environment, including the user’s body, as much as possible to avoid interfering with existing ownership.

\subsection{Illusions in Mediated Reality}
Mediated reality is a concept encompassing the AR of real-world augmentation and the DR of real-world reduction, especially of visual information \cite{mann1999mediated}.
With the shared motivation of VR-driven visuo-haptic research, researchers have investigated cross-modal effects on haptic perception in mediated reality or AR and DR independently.

Visuo-haptic effects are observed in AR spaces with dynamic objects, such as two virtual objects colliding \cite{issartel2015perceiving}, soft objects with virtually dynamic appearances \cite{taima2014controlling}, and virtual content attached to a real moving object \cite{mori2022exploring}.
Static visual effects can also change a haptic sensation. Objects with modified brightness lead to controlled weight, thus reducing muscle fatigue \cite{ban2013augmented}.
Objects with overlaid larger virtual content can induce the SWI \cite{rohrbach2021fooling}. Similarly, such AR visual effects have been explored for object roughness \cite{somada2008psychophysical}, object softness \cite{Punpongsanon15}, and the estimation of COGs \cite{omosako2012shape}.

DR visual effects \cite{mori2017survey} have been less explored than those in AR, specifically in the context of visuo-haptics but discussed in the direction of task performance improvement. Pioneering work by Buchmann et al. demonstrated that semitransparent hands can improve task performance in a fixed viewpoint setup \cite{Buchmann05}.
Similarly, Cosco et al. removed a bulky haptic device occluding a workspace and examined how that helps improve task performance \cite{cosco2012visuo}.
Cheng et al. provide a comprehensive discussion of how different types of DR visual effects influence perceived performance \cite{Cheng22CHI}.

There are only a few works on the impacts of DR visual effects on haptic sensations.
Cosco et al. revealed that the misalignment between a virtual object superimposed on a haptic device, which was obscured by \ac{DR} technology, and the actual position of the haptic device affected the perception of stiffness \cite{cosco2012visuo}.
Hashiguchi et al. explored the weight sensation of partly removed objects \cite{hashiguchi2018perceived}.
Matsumuro et al. used transparent arms to reduce the pain caused by electrical stimulation \cite{matsumuro2022top}.

\begin{figure}
    \centering
    \includegraphics[width=\columnwidth]{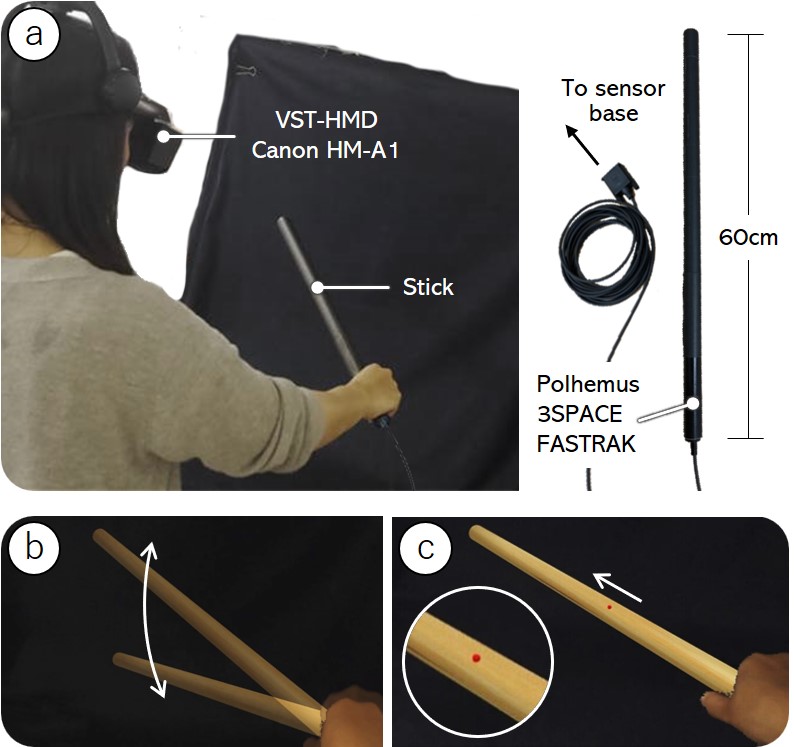}
    \caption{Setup of our study.
    (a) Participants held a $60$ cm stick and observed it through a VST-HMD. Both the HMD and the stick were tracked using Polhemus magnetic tracking sensors.
    (b) The participants uniformly waved the stick to a metronome sound. They verbally reported the perceived weight in magnitude in experiments 1-1 and 1-2.
    (c) The participants reported the COG by shifting the red dot and pressing a keyboard button to indicate the location in experiments 1-2 and 2-2.
    }
    \label{fig:setup}
\end{figure}

We extend the research direction of Hashiguchi et al. \cite{hashiguchi2018perceived} and explore more complex visualizations using DR technology. To the best of our knowledge, we are the first to examine the haptic effects associated with gripped objects from which portions have been partly removed.

\begin{figure}
    \centering
    \includegraphics[width=\columnwidth]{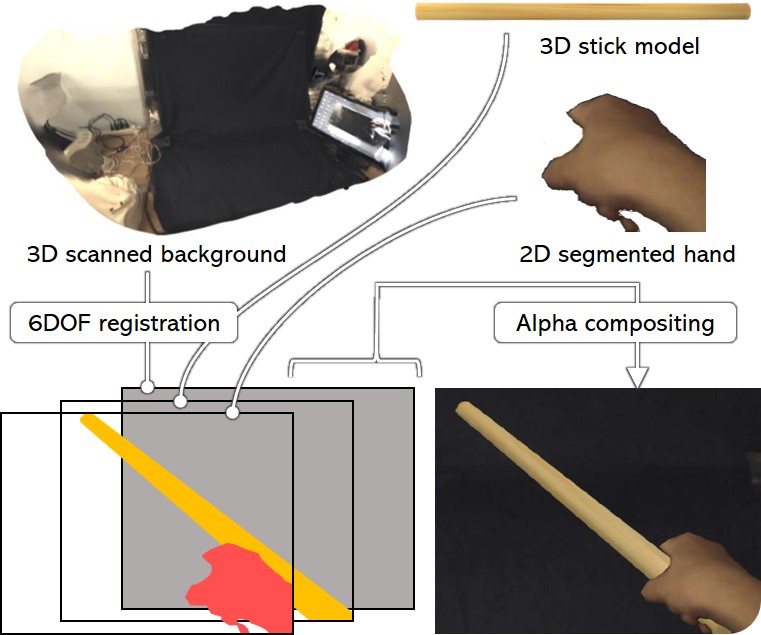}
    \caption{Image processing.
    We prepared a 3D-scanned replica of our study's setup and a 3D model of the stick in advance and registered them in the mediated reality space. We extracted the participant’s hand in the screen space and overlaid it onto the rendered scene and the stick. The alpha composition of all the rendered content resulted in a mediated reality space from a user's perspective. Because the background layer covered the real stick, we were able to render an arbitrarily sized stick.}
    \label{fig:image_proc}
\end{figure}

\begin{figure*}[!t]
    \centering
    \includegraphics[width=\textwidth]{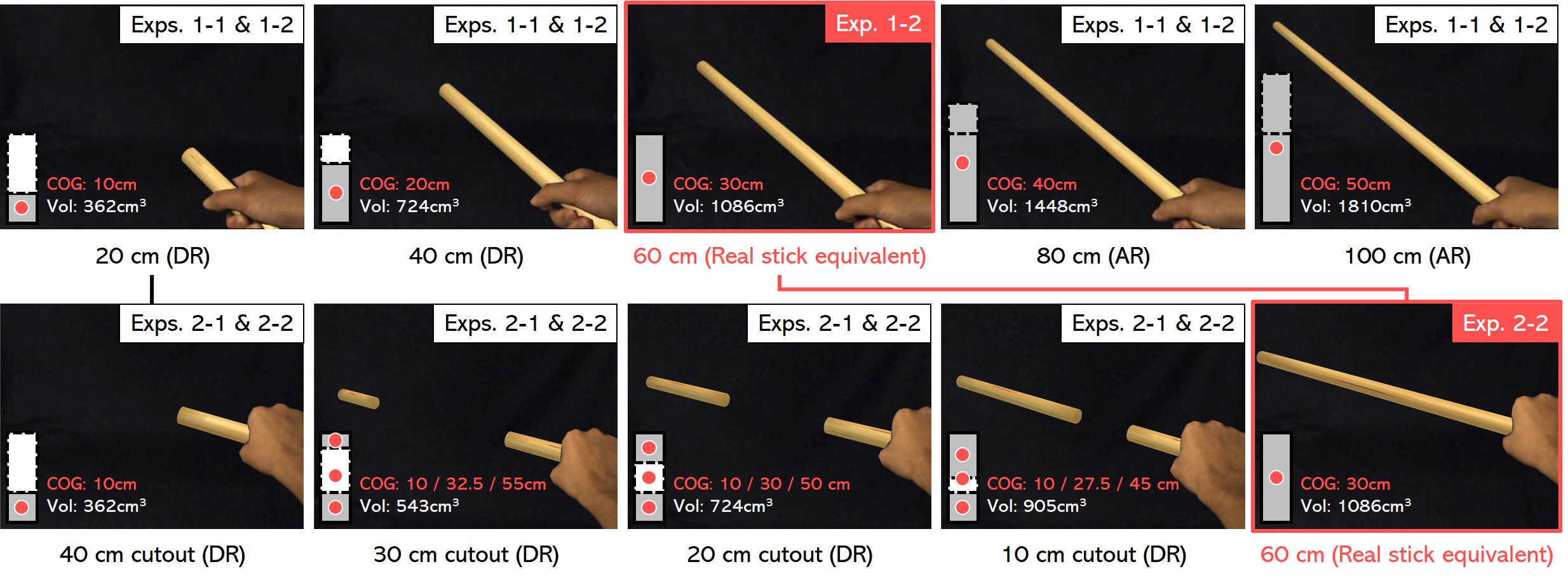}
    \caption{Visual stimuli. The real stick was $60$ cm. Therefore, the conditions of $l = 20$ and $40$ showed virtually shortened sticks while $l = 80$ and $100$ showed virtually extended sticks. In experiments 2-1 and 2-2, we introduced $l_\mathrm{dim} = 10$, $20$, $30$, and $40$ cm, representing sticks that are partially missing their middle portions. Note that the conditions $l=20$ and $l_\mathrm{dim}=40$ are the same. The conditions $l=60$ and $l_\mathrm{dim}=0$ are the same as the real stick. The COG of $l_{\mathrm{dim}}=30$, $20$, and $10$ cm sticks depends on whether users recognize them as a whole or as independent parts.}
    \label{fig:visual_stimuli}
\end{figure*}

\section{Overview}

We summarize our four user-involved experiments to clarify the objectives and rationale behind investigating the perception of mediated reality sticks. We also provide an overview of the common properties shared by the experiments.

\subsection{Study Design}

Our intuition from the findings in SWI led to an assumption that differently sized mediated reality sticks with the same mass would influence the perceived weight of the sticks. However, a research question remains regarding how much the length impacts our perception of weight \cite{thomas2015selection}.
We lend the formulations of dynamic touch to quantitatively relate the perceived weight and physical properties, or inertia. 
To calculate the recognized inertia, we measured perceived COG instead of the length, which is traditionally measured in dynamic touch but visually apparent in our case.

\begin{itemize}
    \item \emph{Experiment 1-1: Weight under Stretched.} We investigated the weight of a virtually extended and shortened stick in mediated reality. The physical mass was constant as the solid maintains the same composition (\figurename~\ref{fig:teaser}, Shortened and Extended). The participants $(N=10)$ were asked to report the perceived weight of such a stick in magnitude while waving it (\figurename~\ref{fig:setup}b).
    \item \emph{Experiment 1-2: \ac{COG} under Stretched.} We analyzed the \ac{COG} locations of the same mediated reality stick from experiment 1-1. The same participants from Experiment 1-1 ($N=10$) also took part in this experiment. The participants reported their  \ac{COG} estimates by marking it with a red dot shifting along the tracked device by pressing a keyboard button to move the position of the red dot (\figurename~\ref{fig:setup}c).
\end{itemize}

The results of experiments 1-1 and 1-2 suggest that the reported weight is closely related to perceived COG locations.
To explain how COG locations can be recognized better, we presented another type of mediated reality stick.
Namely, we visually cut out the middle part to clarify if the relationship between the reported weight and the COG found in the previous experiments would remain (\figurename~\ref{fig:teaser}, Cutout).

The COG may vary depending on whether the sticks recognize the stick as separate parts (i.e., a handle or tip) or as a whole. We speculated that the identified COG would reasonably explain the perceived weight. 

\begin{itemize}
    \item \emph{Experiment 2-1: Weight under Cutout.} We examined the weight of partly missing sticks in mediated reality $(N=10)$. The mass of the sticks in the real world was unchanged while a part appeared to be missing (\figurename~\ref{fig:teaser}, Cutout). The participants reported their estimated weights, as in Experiment 1-1.
    \item \emph{Experiment 2-2: \ac{COG} under Cutout.} We analyzed the COG locations of the same mediated reality stick of Experiment 1-2 $(N=10)$. The participants were the same as those who participated in Experiment 2-1. As in Experiment 1-2, the participants reported their estimates along a stick by marking it.
\end{itemize}

\subsection{Common Properties}
\emph{Apparatus.}
We built a mediated reality system using a Canon MREAL Platform System (MP-110) with a video see-through (VST-) \ac{HMD} (Canon, HM-A1).
We tracked the 6DOF HMD pose using a magnetic sensor running at 120~Hz (Polhemus, 3SPACE FASTRAK).
We chose this VST-HMD because of its camera display axis-aligned optical system for the minimum distortions that could appear in the displays.
The system operated at $30$ fps. Three participants joined a preliminary study investigating the overall system's performance and reported no noticeable delays between real-world events and the displayed stimuli.

We used a stick made from an ABS plastic pipe ($100$ g, $24$ mm diameter, and $60$ cm long) with a small implanted magnetic sensor (Polhemus, Teardrop Mini, RX1-D) for its $6$ DOF tracking (\figurename~\ref{fig:setup}a).
To ensure a uniform tactile feel from grabbing, the stick was wrapped with a smooth material.

\emph{Image processing.}
For stable performance and robust experiments, we took the simplest approach to create a mediated reality environment.
We took photographs covering the laboratory space and created a 3D replica using the photogrammetry software Agisoft Photoscan (\figurename~\ref{fig:image_proc}).
We manually registered the textured 3D mesh to the tracking system and displayed it in the HMD.
We then cropped out the hand area from the captured scene by thresholding the color space using one of the MREAL functions and overlaid the hand area onto the rendered replica.
We registered a 3D stick model onto the tracked stick to virtually extend the stick. We also overlaid the replica of the rendered scene onto the stick to partly remove it. \figurename~\ref{fig:visual_stimuli} summarizes all the visual stimuli investigated in our experiments.

\section{Stretched Mediated Reality Stick}
\label{sec:stretched}
We investigated how stretched mediated reality sticks influence perceived weights and COG. To do this, we performed two experiments to study the relationships between the visually modified lengths and perceived weight (Section \ref{sect:exp1}) and between the visually modified lengths and perceived COG (Section \ref{sect:exp2}). We finally summarized the interpretations from the study results (Section \ref{sect:exp1_2_discussion}).

Ten participants volunteered to join these experiments (all males, $\bar{X}=22.7~(\mathrm{SD}=0.8)$ years old, right-handed). All participants were university students majoring in computer science.

\subsection{Experiment 1-1: Weight under Stretched}
\label{sect:exp1}

\emph{Design.}
We designed a repeated measures within-subjects study to analyze the perceived weight of the mediated reality stick. We introduced an independent variable $l \in \{20, 40, 80, 100\}$ representing the different lengths of the stick. The real stick was 60 cm. Therefore, conditions of $l = 20$ and $40$ were virtually shortened sticks, while $l = 80$ and $100$ were virtually extended sticks. We collected ratings of perceived weight, $w_{l}$, as a dependent variable, the participants verbally reported for a condition of $l$. For example, $w_{40}$ represents the rating for $l=40$.

\emph{Task.}
We performed a magnitude estimation study to collect the participants’ perception of the stick's weight. The participants were instructed to wave a virtual stick as follows: A stick of $l=60$ was presented as having a reference weight of $w_{60}=100$ in magnitude. Then, a randomly selected shorter or longer stick was presented. The participants evaluated the stick and verbally reported the weight in magnitude, $w_l$, compared with that of the reference. For instance, a twice-heavier stick and a twice-lighter stick were $w_l=200$ and $50$, respectively.

\emph{Procedure.}
After completing a consent form and demographic questionnaires, each participant underwent a training session to learn how to wave the stick in a predetermined manner: The participants were asked to maintain a standing posture throughout the experiment. After wearing an HMD, the participants were asked to gently grip the end of the stick at navel height with their right hands. The stick was held parallel to the ground. Their elbows were bent approximately $90$ degrees and contacted their bodies while standing. \figurename~\ref{fig:setup}b illustrates the instructed motion. 
In addition, the participants were instructed to wave the stick around their wrists at approximately $40$ degrees, here comprising an upper and lower arc of $20$ degrees each so that the stick always appeared within the FOV.
Accordingly, the participants swung the stick at a constant rhythm, making it easy to maintain this posture.
The participants practiced waving in time to the
metronome's clicking at $100$ bpm.
The practice session was continued until the participants could perform the specified waving motion.

The main session started after the practice session. The reference stimulus, $l=60$, was presented, and then a randomly selected visual stimulus was presented for evaluation in each trial. The participants waved the stick $10$ times each to the metronome sound. The participants verbally reported the perceived magnitude of
weight for the target stimulus.
This evaluation was repeated for every condition, and the set of evaluations was repeated three times. The reference stimulus and target stimulus were presented on every trial.
With $10$ participants and four comparisons three times each, we collected $120~(= 10 \times 4 \times 3)$ ratings. At the end of the session, we collected comments from the participants. The experiment took approximately $50$ minutes per participant. The participants took sufficient breaks every four trials to eliminate muscle fatigue.

\begin{figure}[!t]
    \centering
    \includegraphics[width=\columnwidth]{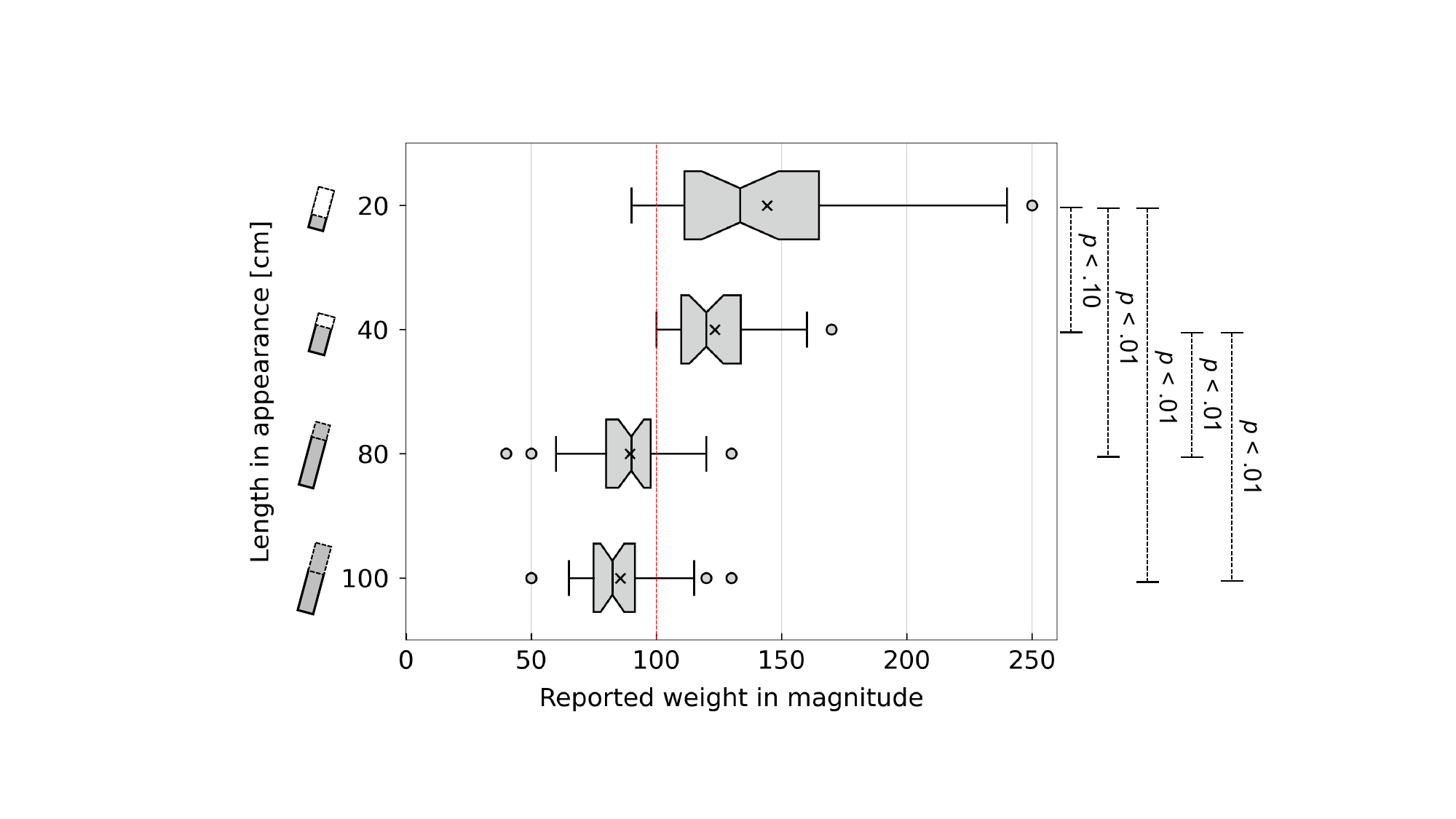}
    \caption{The results of Experiment 1-1. The vertical axis of the graph shows the length of the virtual stick, $l \in \{20, 40, 80, 100\}$, and the horizontal axis shows the magnitude estimate $w_l$. The shorter mediated reality sticks are on the ``heavier'' side ($w_l > 100$), and the longer ones are on the ``lighter'' side ($w_l < 100$). The differences are observed between one of the shorter sticks and one of the longer sticks.}
    \label{fig:exp_1}
\end{figure}

\emph{Results.}
\figurename~\ref{fig:exp_1} summarizes the results of Experiment 1-1.
The Shapiro-Wilk test indicated that the data are consistent with a normal distribution ($p > .05$). 
A one-way repeated-measures analysis of variance (ANOVA) was conducted with $l$ conditions ($l=20, 40, 80, 100$) as the within-subjects factor. Mauchly's test indicated a violation of sphericity ($W=.01, p<.001$). Thus, the Greenhouse-Geisser correction was applied ($\epsilon=.375$). The corrected ANOVA revealed a significant main effect of the $l$ condition ($F(1.13, 10.13)=16.3, p=.002, \eta_{p}^2=.64$, 95\% CI: $[.43, 1.00]$).

The estimated marginal means (EMMs) for each condition were as follows: $l=20$ (EMM = 144.2, SE = 6.97), $l=40$ (EMM = 123.3, SE = 6.97), $l=80$ (EMM = 89.3, SE = 6.97), and $l=100$ (EMM = 85.5, SE = 6.97). Post-hoc pairwise comparisons with a Holm adjustment indicated that \(l=20\) was significantly different from \(l=80\) \((t = 5.569, p < .001, \mathrm{Cohen's}~d = 1.16, 95\%~\mathrm{CI}~[0.33, 1.96])\) and \(l=100\) \((t = 5.958, p < .001, \mathrm{Cohen's}~d = 1.39, 95\%~\mathrm{CI}~[0.49, 2.26])\). Additionally, \(l=40\) was significantly different from \(l=80\) \((t = 3.449, p = .006, \mathrm{Cohen's}~d = 1.30, 95\%~\mathrm{CI}~[0.43, 2.14])\) and \(l=100\) \((t = 3.838, p = .003, \mathrm{Cohen's}~d = 1.77, 95\%~\mathrm{CI}~[0.74, 2.77])\). A marginal trend was observed between \(l=20\) and \(l=40\) \((t = 2.120, p = .087, \mathrm{Cohen's}~d = 0.84, 95\%~\mathrm{CI}~[0.09, 1.55])\). No other pairwise comparisons reached statistical significance \((p > .10)\) (TABLE \ref{tab:stats_posthoc_exp1}).

\begin{table}[!t]
    \centering
    \caption{The statistical results of the pairwise comparisons with Holm adjustments in Experiment 1-1. The values in the table are $p$-values. $(**)$ and $(\dagger)$ indicate statistically significant differences at $p<.01$ and marginal significance at $p<.10$, respectively.}
    \label{tab:stats_posthoc_exp1}
    \begin{tabular}{l|lll}
        \toprule
         & $l=20$ & $l=40$ & $l=80$ \\
        \midrule
        $l=40$ & $0.0867~(\dagger)$ & n/a & n/a \\
        $l=80$ & $<.0001~(**)$ & $0.0056~(**)$ & n/a \\
        $l=100$ & $<.0001~(**)$ & $0.0027~(**)$ & $0.7004$ \\
        \bottomrule
    \end{tabular}
\end{table}

\subsection{Experiment 1-2: COG under Stretched}\label{sect:exp2}
\emph{Design.}
We conducted a repeated measures within-subjects study to evaluate the perceived \ac{COG} of the mediated reality stick. We introduced an independent variable $l \in \{20, 40, 60, 80, 100\}$ of representing the lengths. Notably, $l = 60$, which was a reference in the previous study, was included because COG can be reported without such a reference.
As a dependent variable, we collected the perceived \ac{COG} positions along the stick, $cog_l$, as reported by the participants for $l$.

\emph{Task.}
We undertook an experiment to collect the participants' perceived \ac{COG}. We presented a randomly selected condition, $l$. A red dot was superimposed on the stick at the grasped end ($l = 0$) to allow the participants to report the \ac{COG}'s location, $cog_l$ (\figurename~\ref{fig:setup}c). The participants were asked to locate the dot by shifting it with the up and down arrow keys on a keyboard.

\emph{Procedure.}
After filling out a consent form and demographic questionnaires, each participant was introduced to a training session to learn how to wave the stick. The method of grasping and waving the stick was the same as in Experiment 1-1. After this introductory session, each participant evaluated a randomly selected condition, $l$. The participants waved the stick to the metronome sound at $100$ bpm until they became sure of the \ac{COG}. Upon their reporting, a red dot was superimposed on the stick. The participants indicated the perceived COG by the dot controlled by the keyboard button. Because the stick was held by subjects' dominant right hands, the keyboard was controlled with their left hands. \figurename~\ref{fig:setup}c illustrates the setup. This evaluation was repeated until every condition was assessed, and one set of evaluations was repeated three times. 

With $10$ participants and five conditions at three times each, we collected a total of  $150~(= 10 \times 5 \times 3)$ ratings. At the end of the session, we collected comments from the participants. The experiment took approximately $60$ minutes per participant. The participants took sufficient breaks every five trials to eliminate muscle fatigue.

\begin{figure}[t!]
    \centering
    \includegraphics[width=\columnwidth]{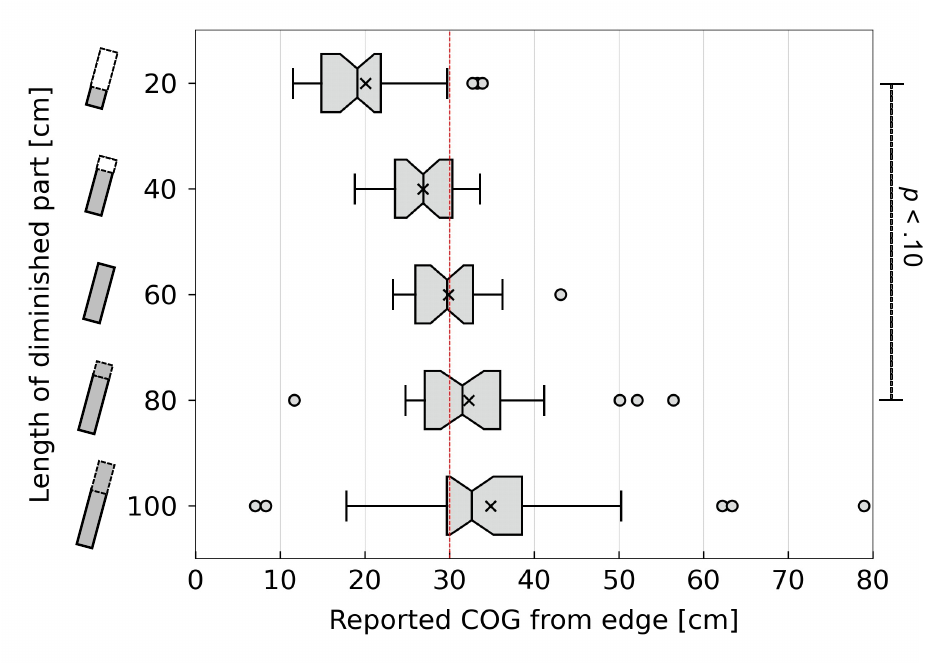}
    \caption{The results of Experiment 1-2. The vertical axis shows the length of the virtual stick presented, $l \in \{20, 40, 60, 80, 100\}$, and the horizontal axis shows the position of the \ac{COG} reported by the participants, $cog_l$. Only a marginal difference was found in $l_{20}$--$l_{80}~(p<.10)$.}
    \label{fig:exp_2}
\end{figure}

\emph{Results.}
\figurename~\ref{fig:exp_2} summarizes the results of Experiment 1-2.
The Shapiro-Wilk test revealed that the data were not normally distributed ($p < .05$). A Friedman's test ($\chi^2 (4) = 22.16, p < .01$) revealed that the difference between the rank sums at each level was significant. Therefore, the length of the virtual stick was detected as affecting the position of the COG.
Pairwise comparisons using the Wilcoxon signed rank test with continuity correction revealed the marginal difference between $20$ cm and $80$ cm (adjusted with Holm $p = .059$, Wilcoxon $V = 0$, $r = -.87$). This result suggests that, for an extended mediated reality stick, the COG position was recognized as being further from the hand than it was, and vice versa.

The median COG values demonstrated tendencies for the COG to shift away from the hand according to the monotonic length. The COG of shorter sticks (i.e., $l \in \{20, 40\}$) was identified as closer to the hand than the physical COG. Similarly, the COG of longer sticks (i.e., $l \in \{80, 100\}$) was identified as further from the hand. However, such effects were subtle and not statistically supported because of the large distributions (TABLE \ref{tab:stats_posthoc_exp2}).

\subsection{Discussion}\label{sect:exp1_2_discussion}

Overall, we found that ``different lengths of a mediated reality stick influence the stick's perceived weight'' in Experiment 1-1.
The results in Experiment 1-1 also support that ``the perceived weight is larger when the length is shorter and vice versa'' but only partially.

Perceived weight linearly increases with the rotational inertia of an observed object and its enhanced speed \cite{gibson1966senses}.
Similarly, we speculated that the weight sensation of our mediated reality sticks can be formulated by inertia. Unlike the conditions described in the literature, our participants assessed the perceived weight while observing sticks with modified apparent lengths. Therefore, the total length of the sticks was apparent, and we collected the COG. Consequently, we attempted to relate the reported COG and perceived weight by calculating inertia based on the COG as understood by the participants.

\begin{figure*}[t!]
    \centering
    \includegraphics[width=\textwidth]{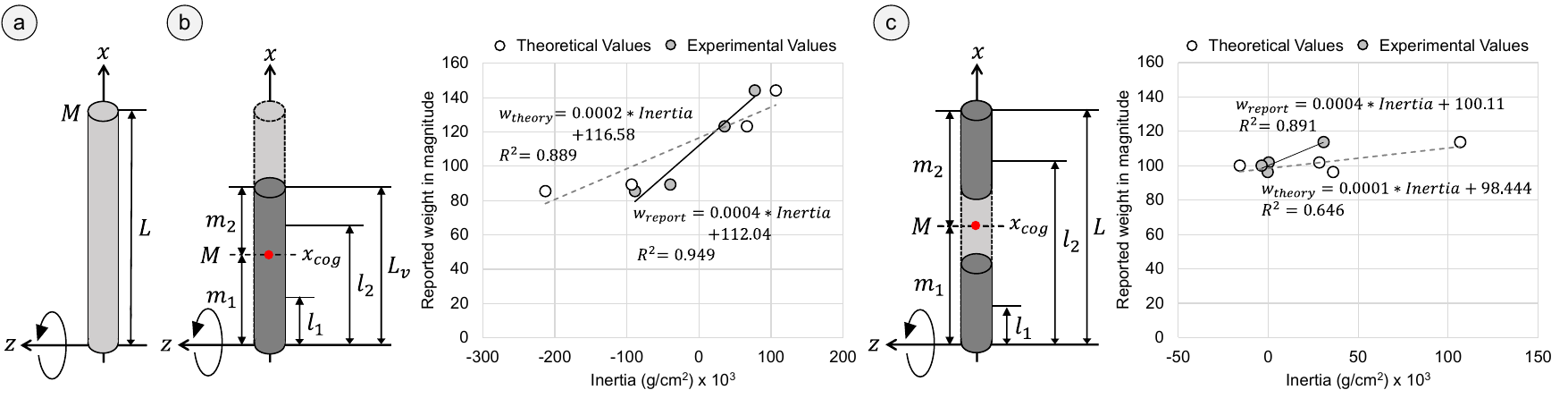}
    \caption{Relationship between the recognized inertia and the reported weight.
    (a) Schematics and parameters for the inertia calculation of the real object.
    (b) Schematics and parameters for the inertia calculations of a mediated reality stick in Experiment 1 based on the reported COG (left) and plots (right).
    (c) Schematics and parameters for the inertia calculations of a mediated reality stick in Experiment 2 based on the reported COG (left) and plots (right).
    Each plot compares inertia with $x_{cog} = L_v / 2$ of the physical stick and the reported COG. The $R^2$ values show that the reported COG fits the reported weights better.
    }
    \label{fig:inertia_real_exp1and2}
\end{figure*}

We calculate the difference of two kinds of inertia, $\Delta I$, as follows:
\begin{equation}
\Delta I = I_r - I_v,
\label{Equation1}
\end{equation}
where $I_r$ is the inertia of the real stick and $I_v$ is the inertia estimated from the appearance of the mediated reality stick (\figurename~\ref{fig:inertia_real_exp1and2}).
We expect our model to capture the discrepancy in inertia between the physical stick and the visually recognized sticks.

The physical stick has its COG in the middle and a uniformly distributed mass, $M$, along its length, $L$.
Therefore,
\begin{equation}
I_r = \int_0^L \frac{M}{L}x^2 dx=\frac{1}{3}ML^2.
\label{Equation2}
\end{equation}
We calculate $I_v$ as follows:
\begin{equation}
I_v = \int_0^{x_{cog}} \frac{m_1}{x_{cog}} x^2 \, dx + \int_{x_{cog}}^{L_v} \frac{m_2}{L_v - x_{cog}} x^2 \, dx.
\label{Equation5}
\end{equation}
The implication is that the stick consists of two parts with different masses, $m_1$ and $m_2$, that add up to $M$ and the two parts are connected at the reported COG location, $x_\text{cog}$ within the apparent length, $L_v$.
Given the individual COG locations of the two masses, $l_1$ and $l_2$, we calculate $x_\text{cog} = (m_1 l_1 + m_2 l_2)/(m_1 + m_2)$.
We substitute $M=m_1+m_2$ for this to calculate the unknown values, $m_1$ and $m_2$.

\begin{table}[!t]
    \centering
    \caption{The statistical results of the Wilcoxon signed rank test in Experiment 1-2. The values in the table are $p$-values, adjusted using the Holm method. $(\dagger)$ for marginal significant different pairs with $p<.10$.}
    \label{tab:stats_posthoc_exp2}
    \begin{tabular}{l|llll}
        \toprule
         & $l=20$ & $l=40$ & $l=60$ & $l=80$ \\
        \midrule
        $l=40$ & $0.130$ & n/a & n/a & n/a \\
        $l=60$ & $0.130$ & $0.399$ & n/a & n/a \\
        $l=80$ & $0.059~(\dagger)$ & $0.399$ & $0.415$ & n/a \\
        $l=100$ & $0.130$ & $0.399$ & $0.399$ & $0.399$ \\
        \bottomrule
    \end{tabular}
\end{table}

We substituted $L_v / 2$ for $x_\text{cog}$ and confirmed that the reported COG as $x_\text{cog}$ fits the collected perceived weights better: $y = 0.0004x + 112.04~(R^2=0.949)$ vs. $y = 0.0002x + 116.58~(R^2=0.889)$, as in \figurename~\ref{fig:inertia_real_exp1and2}b.
The fitting with $x_\text{cog} = L_v / 2$ still reports high $R^2$ values without measuring the recognized COG.
The results indicate that the participants reported a monotonic increase in perceived weight with increasing inertia.
This result corroborates the previous findings regarding dynamic touch \cite{gibson1966senses,loomis1986tactual,streit2007rotational}.

\section{Cutout Mediated Reality Stick}
\label{sec:cut-out}

We observed that the recognized COG played an important role in explaining the perceived weight in Section \ref{sec:stretched}.
For a deeper understanding, we prepared more complex visual stimuli, cutout mediated reality sticks (\figurename~\ref{fig:teaser}, Cutout). mediated reality sticks with cutouts can have different COG locations depending on how participants understand the objects.
The advantage of investigating such visual stimuli is that we can find perceived weights depending either on the areas in sight or the recognized COG, leading to the inertia calculation (\figurename~\ref{fig:visual_stimuli}, second row).

As in the previous section, we conducted two separate experiments for weight (Section~\ref{sect:exp3}) and COG (Section~\ref{sect:exp4}) before drawing an overall conclusion (Section~\ref{sect:exp3_4_discussion}).
Ten participants volunteered to join these experiments (all males, $\bar{X}=22.0~(\mathrm{SD}=0.6)$ years old, right handed). All participants were university students majoring in computer science. The participants were different from those who participated in experiments 1-1 and 1-2.

\subsection{Experiment 2-1: Weight under Cutout}
\label{sect:exp3}

\emph{Design.}
We designed a repeated measures within-subjects study to analyze the perceived weight of the mediated reality stick with a part virtually missing from the middle. We introduced an independent variable $l_\mathrm{dim} \in \{10, 20, 30, 40\}$, here representing the length of the virtually removed portion. The cutout started at $20$ cm, and the $60 - l_\mathrm{dim}$ cm tip remained (\figurename~\ref{fig:visual_stimuli}).
Therefore, $l_\mathrm{dim}=40$ was equivalent to $l=20$ in experiments 1-1 and 1-2. We collected ratings of perceived weight reported by the participants verbally, $w_{l_\mathrm{dim}}$, as a dependent variable.

\emph{Task.}
We performed a magnitude estimation study to evaluate perceived weight. As in Experiment 1-1, each participant was asked to wave the reference stick of $w_0=100$ with no visual removal. Subsequently, they waved a randomly selected stick and verbally reported the weight in magnitude, $w_{l_\mathrm{dim}}$, compared with the reference weight. The participants were not explicitly informed whether the visible parts were one single object. This procedure was repeated until all conditions were covered. This set of evaluations was repeated three times.

\emph{Procedure.}
We followed the procedure of Experiment 1-1. With $10$ participants and four comparisons three times each, we collected a total of $120~(= 10 \times 4 \times 3)$ ratings. The experiment took approximately $50$ minutes per participant.

\begin{figure}[t!]
    \centering
    \includegraphics[width=\columnwidth]{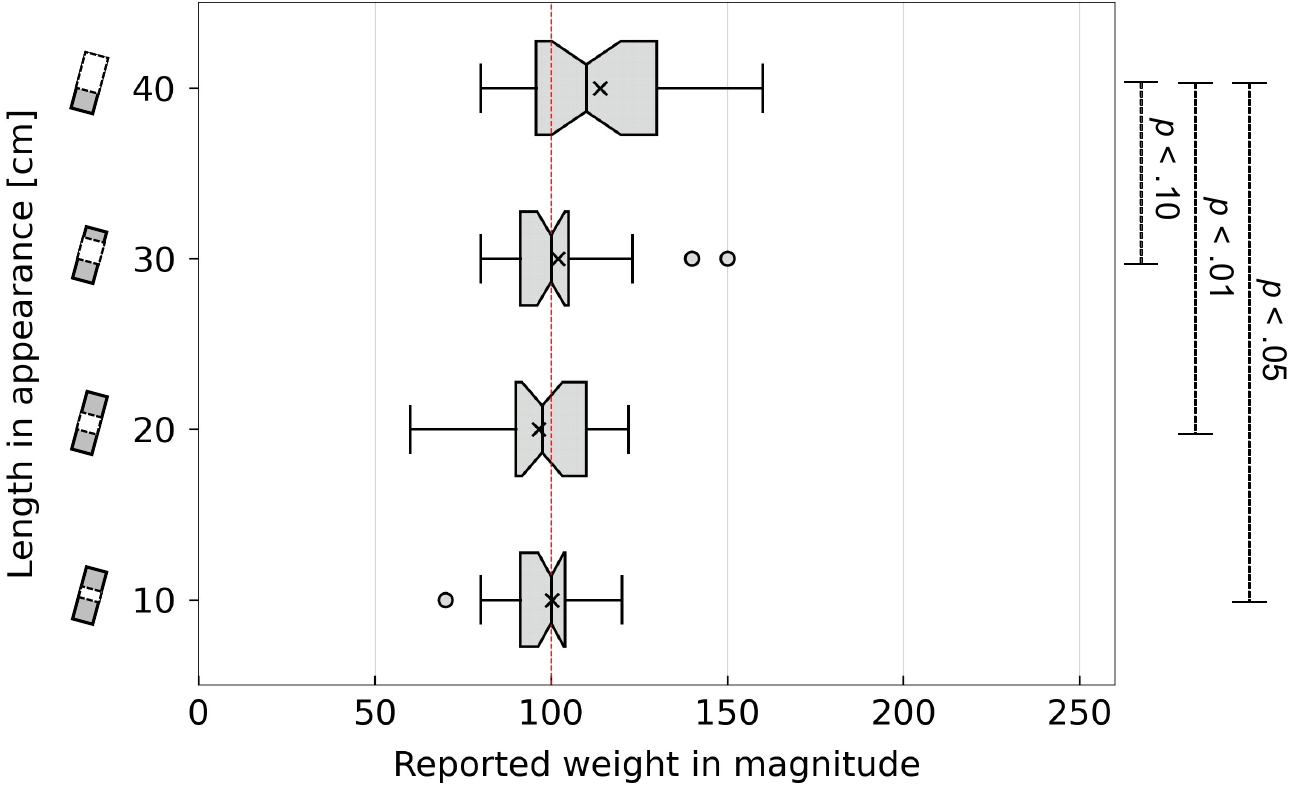}
    \caption{The results of Experiment 2-1. 
    The vertical axis shows the length of the middle part that was removed, $l_\mathrm{dim} \in \{10, 20, 30, 40\}$, and the horizontal axis shows the magnitude of the evaluation value, $w_{l_\mathrm{dim}}$.}
    \label{fig:exp_3}
\end{figure}

\emph{Results.}
The results of Experiment 2-1 are illustrated in \figurename~\ref{fig:exp_3}. 
The Shapiro-Wilk test indicated that the data are consistent with a normal distribution ($p > .05$).
A one-way repeated-measures ANOVA was conducted with the $l_{\mathrm{dim}}$ condition ($l_{\mathrm{dim}}=40, 30, 20, 10$) as the within-subjects factor. Mauchly's test indicated that the assumption of sphericity was met ($W=.26, p=.064$). The ANOVA revealed a significant main effect of the $l_{\mathrm{dim}}$ condition ($F(3,27)=4.86, p=.008, \eta_{p}^2=.35$, 95\% CI: $[.08, 1.00]$).

The EMMs for each condition were as follows: $l_{\mathrm{dim}}= 40$ (EMM = 113.8, SE = 3.4), $l_{\mathrm{dim}}= 30$ (EMM = 101.9, SE = 3.4), $l_{\mathrm{dim}}= 20$ (EMM = 96.4, SE = 3.4), and $l_{\mathrm{dim}}= 10$ (EMM = 100.1, SE = 3.4). 
Post-hoc pairwise comparisons with Holm adjustments indicated that \(l_{\mathrm{dim}}=40\) was significantly different from \(l_{\mathrm{dim}}=20\) \((t = 3.610, p = .007, \mathrm{Cohen's}~d = 0.85, 95\%~\mathrm{CI}~[0.11, 1.57])\) and \(l_{\mathrm{dim}}=10\) \((t = 2.834, p = .043, \mathrm{Cohen's}~d = 0.74, 95\%~\mathrm{CI}~[0.02, 1.43])\). A marginal trend was observed between \(l_{\mathrm{dim}}=40\) and \(l_{\mathrm{dim}}=30\) \((t = 2.460, p = .082, \mathrm{Cohen's}~d = 0.85, 95\%~\mathrm{CI}~[0.10, 1.56])\). No other pairwise comparisons reached statistical significance \((p > .10)\) (TABLE \ref{tab:stats_posthoc_exp3}).

\subsection{Experiment 2-2: COG under Cutout}
\label{sect:exp4}
\emph{Design.}
We conducted a repeated measures within-subjects study to analyze the perceived \ac{COG} of the cutout stick. We introduced an independent variable, $l_\mathrm{dim} \in \{0, 10, 20, 30, 40\}$, here representing the length of the virtually removed portion. What was different from Experiment 2-1 was $l_\mathrm{dim} = 0$, which was equivalent to the original stick.
As a dependent variable, we collected the perceived \ac{COG} locations, $cog_\mathrm{pos}$, as indicated by the up and down keys on a keyboard.

\emph{Task.}
Much like Experiment 1-2, each participant was asked to wave the stick, which was randomly selected, and then manually report the perceived centroid position via keyboard. 
The participants were allowed to report COG locations where the virtual stick was missing.
We presented one of the randomly selected conditions. A red dot was superimposed on the stick at the grasped end ($l = 0$) to allow the participants to report the \ac{COG}'s location, $cog_\mathrm{pos}$.
This evaluation was repeated until all the conditions were evaluated. 

\emph{Procedure.}
We followed the procedure of Experiment 1-2. With $10$ participants and five conditions three times each, we collected a total of $150~(= 10 \times 5 \times 3)$ ratings. The experiment took approximately $60$ minutes per participant.

\emph{Results.}
\figurename~\ref{fig:exp_4} reveals the results. 
The Shapiro-Wilk test indicated that the data are consistent with a normal distribution ($p > .05$). 
A one-way repeated-measures ANOVA was conducted with the $l_\mathrm{dim}$ condition ($l_\mathrm{dim}= 40, 30, 20, 10, 0$) as the within-subjects factor. Mauchly's test confirmed sphericity ($W= .30$, $p = .46$). The ANOVA revealed a significant main effect of the $l_\mathrm{dim}$ condition ($F(4, 36) = 7.21$, $p < .001$, $\eta_p^2 = .44$, 95\% CI: $[.20, 1.00]$).

The EMMs for each condition were as follows: $l_{\mathrm{dim}}= 40$ (EMM = 24.4, SE = .929), $l_{\mathrm{dim}}= 30$ (EMM = 29.8, SE = .929), $l_{\mathrm{dim}}= 20$ (EMM = 29.9, SE = .929), $l_{\mathrm{dim}}= 10$ (EMM = 30.5, SE = .929), and $l_{\mathrm{dim}}= 0$ (EMM = 29.3, SE = .929). Post-hoc pairwise comparisons with Holm adjustments indicated that $l_{\mathrm{dim}}=40$ was significantly different from all other conditions: $l_{\mathrm{dim}}=30$ ($t = -4.120$, $p = .002$, Cohen's $d = -1.02$, 95\% CI $[-1.78, -0.23]$), $l_{\mathrm{dim}}=20$ ($t = -4.191$, $p = .002$, Cohen's $d = -1.14$, 95\% CI $[-1.94, -0.32]$), $l_{\mathrm{dim}}=10$ ($t = -4.665$, $p < .001$, Cohen's $d = -1.33$, 95\% CI $[-2.18, -0.45]$), and $l_{\mathrm{dim}}=0$ ($t = -3.749$, $p = .004$, Cohen's $d = -1.07$, 95\% CI $[-1.84, -0.27]$). No other pairwise comparisons reached statistical significance ($p > .10$) (TABLE \ref{tab:stats_posthoc_exp4}).

\begin{table}[!t]
    \centering
    \caption{The statistical results of the pairwise comparisons with Holm adjustments in Experiment 2-1. The values in the table are $p$-values. $(**)$, $(*)$, and $(\dagger)$ indicate significantly different pairs at $p<.01$, $p<.05$, and $p<.10$, respectively.}
    \label{tab:stats_posthoc_exp3}
    \begin{tabular}{l|lll}
        \toprule
         & $l_\mathrm{dim} =10$ & $l_\mathrm{dim} =20$ & $l_\mathrm{dim} =30$ \\
        \midrule
        $l_\mathrm{dim}=20$ & $0.8888$ & n/a & n/a \\
        $l_\mathrm{dim}=30$ & $0.8888$ & $0.7803$ & n/a \\
        $l_\mathrm{dim}=40$ & $0.0430~(*)$ & $0.0074~(**)$ & $0.0823~(\dagger)$ \\
        \bottomrule
    \end{tabular}
\end{table}

\subsection{Discussion}\label{sect:exp3_4_discussion}
\begin{figure}[t!]
    \centering
    \includegraphics[width=\columnwidth]{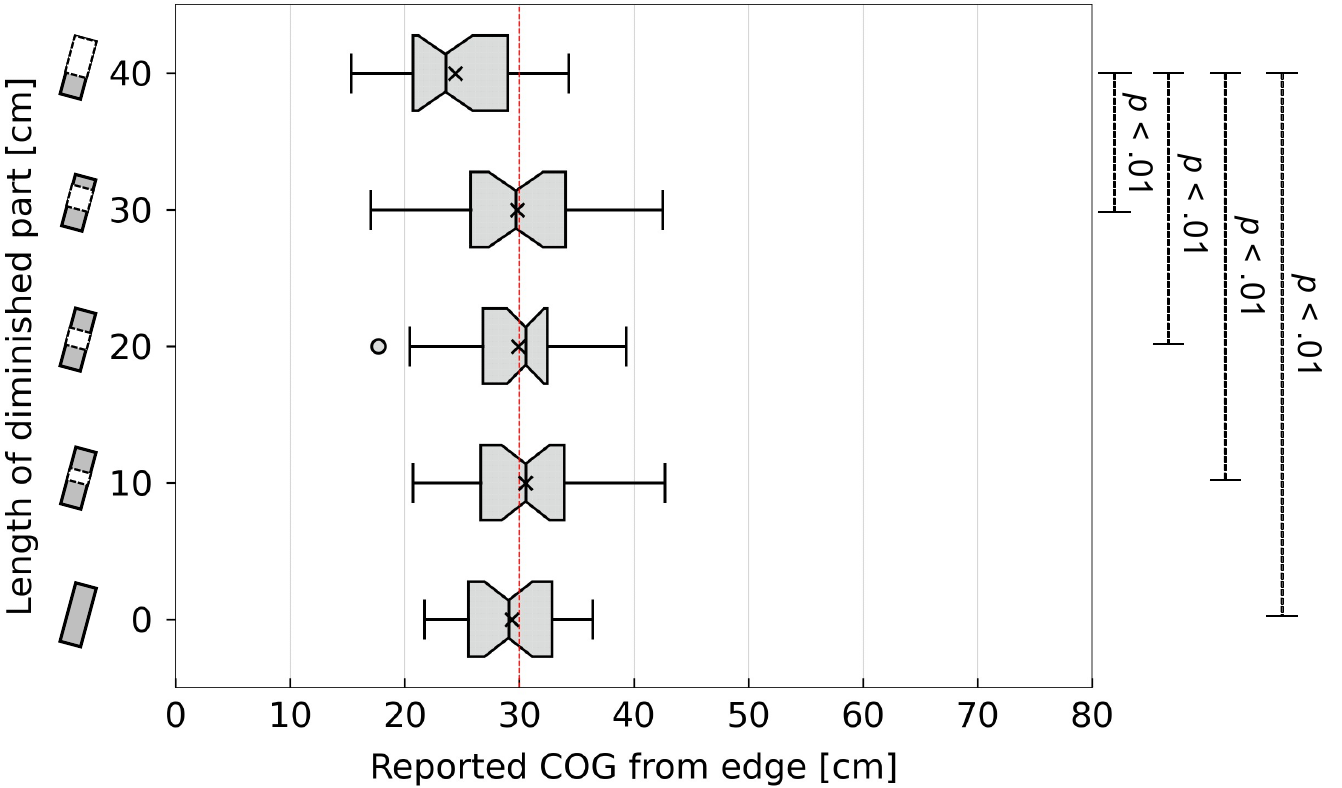}
    \caption{The results of Experiment 2-2. The vertical axis shows the length of the middle portion removed ($l_\mathrm{dim} \in \{0, 10, 20, 30, 40\}$), and the horizontal axis shows the position of the \ac{COG} as reported by the subjects ($cog_\mathrm{pos}$).}
    \label{fig:exp_4}
\end{figure}

Some significant differences were observed across conditions for both reported weight and COG (Figs.~\ref{fig:exp_3} and \ref{fig:exp_4}). However, no statistically significant differences were found for either reported weight or COG specifically among the $l_\mathrm{dim} = 30$, $20$, and $10$ conditions.
According to the reported COG of the diminished part, the participants perceived the stick as rigidly connected, and no participants wished to specify more than one COG. 

We took the same approach as in Experiment 1-2 to calculate inertia (\figurename~\ref{fig:inertia_real_exp1and2}c).
Similarly, we compared the inertia based on the physical COG ($x_\text{cog}=L_v/2$) and the reported COG, and the latter fits the perceived weights better: $y = 0.0001x + 98.444~(R^2=0.646)$ vs. $y = 0.0004x + 100.11~(R^2=0.891)$.

\section{Interpretations and Future Work}

In what follows, we discuss how the mediated reality sticks cause illusionary weights and future extensions.

\subsection{Overall Interpretations}
Our results demonstrate that the discrepancy between the inertia of the physical stick and that of the recognized COG explains the perceived weights.

\begin{table}[!t]
    \centering
    \caption{The statistical results of the pairwise comparisons with Holm adjustments in Experiment 2-2. The values in the table are $p$-values. $(**)$ indicates a statistically significant difference with $p<.01$.}
    \label{tab:stats_posthoc_exp4}
    \begin{tabular}{l|llll}
        \toprule
         & $l_\mathrm{dim} =0$ & $l_\mathrm{dim} =10$ & $l_\mathrm{dim} =20$ & $l_\mathrm{dim} =30$ \\
        \midrule
        $l_\mathrm{dim} =10$ & $1.0000$ & n/a & n/a & n/a \\
        $l_\mathrm{dim} =20$ & $1.0000$ & $1.0000$ & n/a & n/a \\
        $l_\mathrm{dim} =30$ & $1.0000$ & $1.0000$ & $1.0000$ & n/a \\
        $l_\mathrm{dim} =40$ & $0.0044~(**)$ & $0.0004~(**)$ & $0.0015~(**)$ & $0.0017~(**)$ \\
        \bottomrule
    \end{tabular}
\end{table}

In Experiment 1, we confirmed that the stretched stick was perceived to be lighter, while the shorter one was perceived to be heavier than the stretched sticks. 
We consider that the motion commands for longer sticks, which anticipate the perceived COG to be greater than $30$ cm but actually be at $30$ cm, unexpectedly and unnecessarily induce a stronger arm motion, resulting in a lighter feeling. Conversely, motion commands for shorter sticks, which anticipate a perceived COG of less than 30 cm, unexpectedly induce weaker arm motion, leading to a heavier feeling.
Therefore, appearance inertia, which is based on predictions and feedback, may affect the inertia of the real stick.
The relationship between visual information such as predictions and feedback and actual tactile information shows a similar tendency to SWI\cite{muller1889, woodworth1924psychology}.
Considering these findings, we have formulated the relationship between inertia and weight based on the findings of dynamic touch, a phenomenon similar to SWI.

In Experiment 2, the perceived weight stayed unchanged in cutout mediated reality sticks, even though the visible area changed.
As depicted in \figurename~\ref{fig:visual_stimuli}, $l=40$ and $l_\mathrm{dim}=20$ have the same apparent area.
Moreover, $l_\mathrm{dim}=30$ is $10$ less than $l=40$, and $l_\mathrm{dim}=10$ is $10$ less than $l=60$. However, the reduced visible area rarely changes the perceived weight.
This result does not follow the SWI, strengthening the support for the relationship between the recognized inertia and weights.
Four of the ten participants commented that ``they felt no difference in weights between the conditions of the real stick and cutout stick,'' which supports our finding.
This result shows that the perceived weight is based on the inertia of the object rather than the visible area.
The fact that the COG locations were reported in the missing areas indicates that the cutout mediated reality sticks were recognized as a single object.

\subsection{Future Work}
\emph{Generalization.}
We have focused on the stick-shaped object. Therefore, how to generalize our findings merits debate. We speculate that the recognized inertia is the explanatory factor for the weights estimated by the participants. There is certain evidence that the weights of differently shaped objects with the same volume will be perceived differently \cite{vicovaro2019influence}. However, it is uncertain if this applies to other cases, such as spheres, triangles, and dynamic objects similar to elastic objects \cite{burton1990can, harris2023mass}.

\emph{Muscle Activity Analysis.}
Direct measurements such as muscle activity analysis would provide more solid clues on how people behave toward visual stimuli, unexpected feedback, and reactions to them \cite{mangalam2019muscular, waddell2016perceived}. Such analysis would also better explain fatigue and, thus, tiredness for long-term usage and how these factors differ between mediated reality and the normal visual world.

\emph{Diversity in Experiences.}
Because the recognized inertia would matter when estimating the weight, we anticipate how well one is trained or has the experience to infer the recognized inertia will influence the weight estimation. Our studies were performed solely with male university students, so there is no denying that our results would have been biased by their capabilities and experiences. Therefore, collecting participants from more diverse groups, especially those of different ages and genders, would be valuable.

\section{Conclusion}
We investigated how the perceived weight and COG of the mediated reality stick change when it is virtually stretched and has a cutout. Through a series of user studies, we discovered that (1) the different lengths of a mediated reality stick influence the participants’ perceived weight, (2) the perceived weight is greater when the length is shorter and vice versa, and (3) the perceived weight and COG of the stick do not change in the cutout. We further discussed how the inertia calculated from the reported COG can explain the perceived weight.
Overall, perceived weights varied with the mediated reality's visual modifications, which show similar effects to SWI and can be related to inertia based on the recognized COG. As such, the phenomenon is safely placed within the context of dynamic touch.

Although there are challenges to generalization, our case study of mediated reality sticks is a milestone in exploring different shapes and conditions. Future extensions of this work would include shapes other than sticks, bent or elastic sticks, muscle activity analysis, long-term analysis, and analysis in diverse groups.

\ifCLASSOPTIONcompsoc
  \section*{Acknowledgments}
\else
  \section*{Acknowledgment}
\fi

This work was partly supported by JSPS KAKENHI Grant Numbers 21K11947 and
the Deutsche Forschungsgemeinschaft (DFG, German Research Foundation) under Germany´s Excellence Strategy – EXC 2120/1 – 390831618.
The authors also thank Ms. Miho Tanaka and Mr. Hiroki Sakiyama for their support in setting up the experiments.

\ifCLASSOPTIONcaptionsoff
  \newpage
\fi

\bibliographystyle{IEEEtran}
\bibliography{references.bib}
%
\vspace{-1.5\baselineskip}
\begin{IEEEbiography}[{\includegraphics[width=1in,height=1.25in,clip,keepaspectratio]{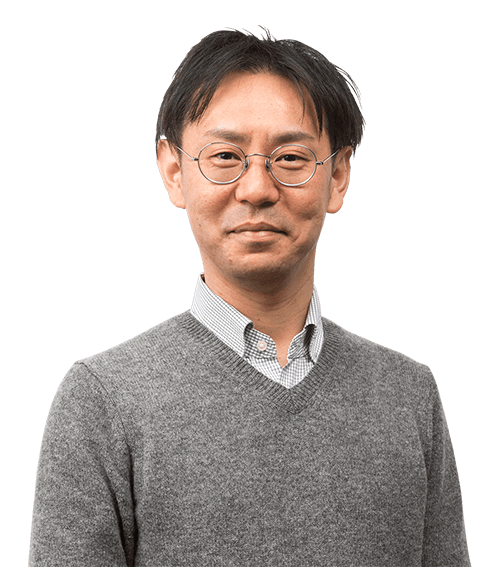}}]{Satoshi Hashiguchi}
is a senior researcher at Ritsumeikan University. He received a Ph.D. in engineering from Kyushu University. His research interests are multi-/cross-modal computing in augmented/virtual reality. 
\end{IEEEbiography}

\vspace{-1.5\baselineskip}
\begin{IEEEbiography}[{\includegraphics[width=1in,height=1.25in,clip,keepaspectratio]{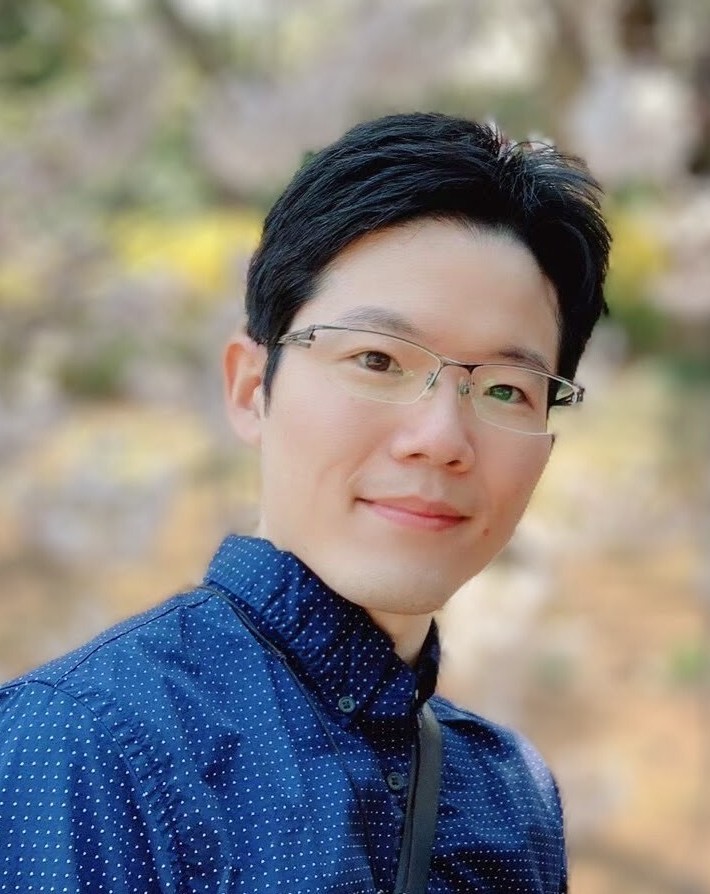}}]{Yuta Kataoka}
is a visiting researcher at Ritsumeikan University. He received his Ph.D. in engineering from Ritsumeikan University, Japan, in 2021. His research interests include human--computer interaction and human perception.
\end{IEEEbiography}

\vspace{-1.5\baselineskip}

\begin{IEEEbiography}[{\includegraphics[width=1in,height=1.25in,clip,keepaspectratio]{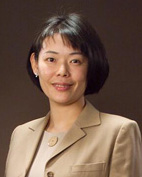}}]{Asako Kimura}
is a professor at Ritsumeikan University. She received her Ph.D. in engineering science from Osaka University, Osaka, Japan, in 2001. She worked at Osaka University from 2000 to 2003 before coming to Ritsumeikan University. She was concurrently a special project associate at the Biomedical Imaging Resource Laboratory, Mayo Clinic, from 2001 to 2002, and a Sakigake Researcher at PRESTO Japan Science and Technology Agency from 2006 to 2010. Her research interests include human-computer interaction, human perception, augmented/mixed reality, and virtual reality. She is a member of the Human Interface Society, the Virtual Reality Society of Japan, IPSJ, IEICE, IEEE, and ACM.
\end{IEEEbiography}

\vspace{-1.5\baselineskip}

\begin{IEEEbiography}[{\includegraphics[width=1in,height=1.25in,clip,keepaspectratio]{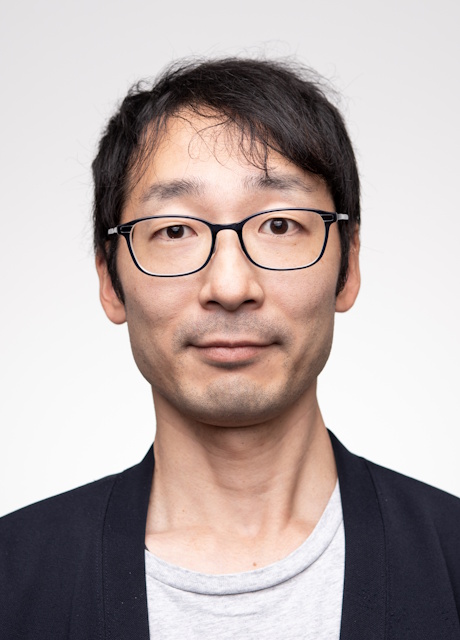}}]{Shohei Mori}
is a Junior Research Group Leader at the Visualization Research Center (VISUS), Cluster of Excellence IntCDC, University of Stuttgart, Germany, and serves as a Guest Associate Professor (Global) at Keio University, Japan. His research focuses on Mediated Reality and its enabling core technologies, including computational imaging and real-time graphics. He received his B.Eng. (2011), M.Eng. (2013), and D.Eng. (2016) from Ritsumeikan University, Japan. He was a Research Fellow for Young Scientists (DC-1 \& PD) under the Japan Society for the Promotion of Science (JSPS), and a postdoctoral researcher at Keio University (2016--2018) and Graz University of Technology (2018--2024). He has received Best Paper and Demo Awards at premier conferences such as IEEE VR and IEEE ISMAR.
\end{IEEEbiography}




\end{document}